\documentclass[10pt,journal,compsoc,final]{IEEEtran}
\ifCLASSINFOpdf
\else
\fi
\usepackage{graphicx,subfigure,subfig,color}
\usepackage{bm,amsmath,amssymb,amsthm,dsfont}
\usepackage{algorithm,algorithmic}
\usepackage{url}
\usepackage{balance}
\usepackage{threeparttable}

\begin{document}
\title{Network-regularized Sparse Logistic Regression Models for Clinical Risk Prediction and Biomarker Discovery}
\author{Wenwen Min, Juan Liu, and Shihua Zhang
\IEEEcompsocitemizethanks{\IEEEcompsocthanksitem Wenwen Min, and Juan Liu are with the State Key Laboratory of Software Engineering, School of Computer, Wuhan University, Wuhan 430072, China. E-mail: \{mww, liujuan\}@whu.edu.cn.

\IEEEcompsocthanksitem Shihua Zhang is with National Center for Mathematics and Interdisciplinary Sciences, Academy of Mathematics and Systems Science, Chinese Academy of Sciences, Beijing 100190, China. E-mail: zsh@amss.ac.cn.}
\thanks{Manuscript received XXX, 2016; revised XXX, 2016.}}
% The paper headers
%\markboth{IEEE/ACM TRANSACTIONS ON COMPUTATIONAL BIOLOGY AND BIOINFORMATICS,~Vol.~XX, No.~XX, XX~2016}
%{Min \MakeLowercase{\textit{et al.}}: Network-regularized Sparse Logistic Regression Models for Clinical Risk Prediction and Biomarker Discovery}

\IEEEtitleabstractindextext{
\begin{abstract}
Molecular profiling data (e.g., gene expression) has been used for clinical risk prediction and biomarker discovery. However, it is necessary to integrate other prior knowledge like biological pathways or gene interaction networks to improve the predictive ability and biological interpretability of biomarkers. Here, we first introduce a general regularized Logistic Regression (LR) framework with regularized term $\lambda \|\bm{w}\|_1 + \eta\bm{w}^T\bm{M}\bm{w}$, which can reduce to different penalties, including Lasso, elastic net, and network-regularized terms with different $\bm{M}$. This framework can be easily solved in a unified manner by a cyclic coordinate descent algorithm which can avoid inverse matrix operation and accelerate the computing speed. However, if those estimated $\bm{w}_i$ and $\bm{w}_j$ have opposite signs, then the traditional network-regularized penalty may not perform well. To address it, we introduce a novel network-regularized sparse LR model with a new penalty $\lambda \|\bm{w}\|_1 + \eta|\bm{w}|^T\bm{M}|\bm{w}|$ to consider the difference between the absolute values of the coefficients. And we develop two efficient algorithms to solve it. Finally, we test our methods and compare them with the related ones using simulated and real data to show their efficiency.
\end{abstract}
}
\maketitle
\IEEEpeerreviewmaketitle
\section{Introduction}
\IEEEPARstart{P}{redicting} clinical risk and discovering molecular prognostic signatures are key topics in genomic studies \cite{liu2014breast,yang2016comparative}. Large number of genomic datasets (e.g., TCGA \cite{ Benz2013The,Hoadley2014Multiplatform}) have been rapidly generated on cancer and other complex disease \cite{Benz2013The,kristensen2014principles, zhao2012efficient, zhang2012discovery}. These available cancer genomic data provides us an unprecedented opportunity to predict the development risk of cancer via integrating diverse molecular profiling data \cite{Zhang2011A,kristensen2014principles}. Recently, some prognostic models have been proposed via integrating clinical and gene expression data \cite{pittman2004integrated,Zhang2013Network,Simon2011Regularization}. Most of these methods are based on the Cox proportional hazards model \cite{Cox1972} and a few are designed based on other machine learning methods for this task. We can also easily dichotomize the survival time into a binary outcome and obtain a typical classification problem which can been solved by many supervised machine learning methods.

Logistic regression (LR) is one of such a classical method and has been widely used for classification \cite{liao2007logistic}. However the traditional LR model employs all (or most) variables for predicting and lead to a non-sparse solution with limited interpretability. The sparsity principle is an important strategy for interpretable analysis in statistics and machine learning. Recently, a number of studies have focused on developing regularized or penalized LR models to encourage sparse solutions and use a limited number of variables for predicting \cite{Shevade2003A,Cawley2006Gene,Wu2009Genome,Bootkrajang2013Classification,Tan2013Minimax,lin2013advanced,Park2016A}. Many of the penalized LR models apply Lasso as a penalty function to induce sparsity. However the Lasso fails to select strongly correlated variables together and tends to select a variable from them \cite{algamal2015regularized}. Thus, many generalizations of the Lasso have been proposed to solve the limits of Lasso \cite{Tibshirani2011Regression}, including elastic net, group Lasso, and fused Lasso, etc. On the other hand, some researchers proposed refined regression models with network-based penalties \cite{li2008network,li2010variable,Zhe2013Joint}. These models are expect to get more accurate prediction and better interpretability via integrating prior knowledge.

Motivated by the development of sparse coding and network-regularized norm \cite{li2008network,cai2011graph}, we address the double tasks -- feature selection and class prediction -- by using a network-based penalty in the Logistic Regression (LR) framework. Specifically, we first focus on a traditional network-regularized penalty:
\begin{equation}
 \mathcal{R}_1(\bm{w}) = \lambda \|\bm{w}\|_1 + \eta\bm{w}^T\bm{L}\bm{w},
\end{equation}
where $\bm{L}$ is the normalized Laplacian matrix encoding a prior network (e.g., a protein interaction network). The first term is a $L_1$-norm penalty to induce sparsity. The second term $\bm{w}^T\bm{L}\bm{w} = \sum_{i\sim j}\bm{A}_{ij}(\bm{w}_i/\sqrt{d_i}-\bm{w}_j/\sqrt{d_j})^2$ is a quadratic-Laplacian norm penalty to force the coefficients of $\bm{w}$ to be smooth. More importantly, it is a convex penalty. Thus, such a regularized LR model and its reduced forms can be solved effectively. However if those estimated $\bm{w}_i$ and $\bm{w}_j$ have opposite signs, then the traditional network-regularized penalty may not perform well. To address this limitation, we focus on a novel network-based penalty \cite{min2016l0}:
\begin{equation}
 \mathcal{R}_2(\bm{w}) = \lambda \|\bm{w}\|_1 + \eta|\bm{w}|^T\bm{L}|\bm{w}|,
\end{equation}
where $|\bm{w}|^T\bm{L}|\bm{w}| = \sum_{i\sim j}\bm{A}_{ij}(|\bm{w}_i|/\sqrt{d_i}-|\bm{w}_j|/\sqrt{d_j})^2$ which is adopted to eliminate the effects of symbols of the estimated coefficients. However, the novel penalty is not differentiable at the zero point. Intuitively, it is a challenge issue to solve such a regularized LR model using the conventional gradient descent method. Here we develop two methods to solve it. We first propose to use the following penalty:
$$ \lambda \sum\limits_{i=1}^p |w_i| + \eta \sum_{i\sim j}\bm{A}_{ij} \left(\frac{\mbox{sign}(\widehat{\bm{w}}_i)\bm{w}_i}{\sqrt{d_i}}-\frac{\mbox{sign}(\widehat{\bm{w}}_j)\bm{w}_j}{\sqrt{d_j}} \right)^2$$
to replace $\mathcal{R}_2$, i.e., $|\bm{w}_i| \approx \mbox{sign}(\widehat{\bm{w}}_i)\bm{w}_i$, where $\widehat{\bm{w}}$ is computed using the maximum likelihood estimation for classical LR model. The new penalty is a convex function and has been used for a regression model \cite{li2010variable}. Similarly, we can effectively solve this regularized LR model. We also try to solve the novel penalty ($\mathcal{R}_2(\bm{w})$) directly. Fortunately, Hastie \emph{et~al.} \cite{hastie2015statistical} (see Page 112) find that $\mathcal{R}_2$ has a good property (i.e., condition regularity \cite{hastie2015statistical,Tseng2001Convergence}), which inspires us to solve it via a cycle coordinate descent algorithm.

To sum up, our key contributions are two-fold. First, we introduce a unified Network-regularized Sparse LR (NSLR) framework enabling the Lasso, elastic net and network-regularized LR models are all special cases of it. More importantly, this framework can be efficiently solved using a coordinate-wise Newton algorithm. Second, we propose a novel network-regularized LR model to eliminate the sensitivity of the conventional network-regularized penalty to the signs of estimated coefficients. Here we adopt an approximate algorithm and a coordinate-wise Newton algorithm to solve it, respectively. Finally, we apply our methods to Glioblastoma multiforme (GBM) gene expression and cancer clinical data from TCGA database \cite{Benz2013The}, and a protein interaction network data from Pathway Commons \cite{Cerami2010Pathway} for survival risk prediction and biomarker discovery. Our first key is to identify some gene biomarkers for survival risk prediction. In addition, based on the prediction probability scores of GBM patients, we can divide them into different subtypes relating to survival output. Furthermore, we apply our methods to a lung cancer dataset with two subtypes including Lung adenocarcinoma (LUAD) and Lung squamous-cell carcinoma (LUSC) to identify subtype-specific biomarkers.
\section{Method}
\begin{figure*}[htbp]
  \centering
  \includegraphics[width = 1\linewidth]{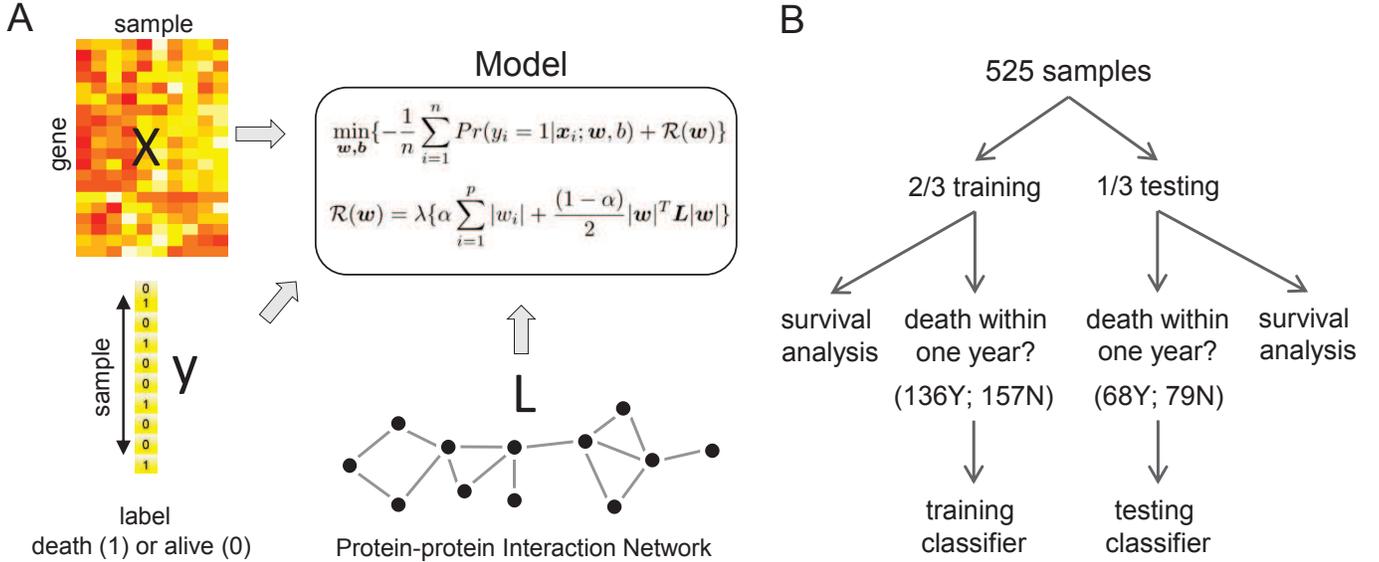}\\
  \caption{(A) A novel network-regularized sparse logistic regression framework by considering the difference between the absolute values of the coefficients. It combines the gene expression data, the normalized Laplacian matrix $\bm{L}$ encoding the protein interaction network and the clinical binary outcome to train a prediction model. (B) Illustration of sample divisions with an application using GBM data from TCGA. Here we collect a total of 525 GBM samples which are randomly divided into two subsets, including 350 training samples and 175 testing samples. We dichotomize the survival time of patients into clinical binary outcomes to formulate it into a classification problem. For each patient, we dichotomize its survival time into a binary outcome based on his/her survival time. For training samples, there are 57 censored patients kept for using in survival analysis. For testing samples, there are 32 censored patients. In the training process, we first remove those censored samples to train the regularized LR models. Then, we re-test all the training data and all the testing data (including censored samples) with the trained models. Based on the predicted LR scores computed via $Pr(y=1|\bm{x};\bm{w})$, we divide all the training and testing samples into different risk groups for further survival analysis.}
  \label{fig:01}
\end{figure*}

In this section, we first briefly review the typical logistic regression model and an iteratively re-weight least squares (IRLS) learning strategy based on a coordinate-wise Newton method. We further introduce the two network-regularized LR models together with the basic regularized LR models (Lasso and elastic net) in a unified framework in Section 2.3 and suggest a general coordinate-wise Newton algorithm for solving this framework. Lastly, we propose a novel network-regularized sparse LR model in Section 2.5. We can use an approximate strategy (AdaNet.LR in Section 2.3) to solve it. In addition, we also develop an efficient coordinate-wise Newton algorithm to solve the AdaNet.LR directly (Section 2.5).
\subsection{Logistic Regression}
Here we consider a binary classification problem. Given $n$ training samples $\mathcal{D} = \{(\bm{x}_1, y_1),\cdots, (\bm{x}_n, y_n)\}$ where $\bm{x}_i \in \mathbb{R}^{p}$ is a $p$-dimensional column vector and label $y_i \in \{0,1\}$. We first write the logistic function as follows:
$$p(\bm{x};\bm{w},b) = Pr(y=1|\bm{x};\bm{w},b) = \frac{1}{1+\exp(-\bm{w}^T\bm{x}-b)}.$$
For convenience, let $\bm{w} = (\bm{w},b)$ and $\bm{x} = (\bm{x},1)$. We can rewrite it:
\begin{equation}\label{eq-3}
  p(\bm{x};\bm{w}) = Pr(y=1|\bm{x};\bm{w}) = \frac{1}{1+\exp(-\bm{w}^T\bm{x})},
\end{equation}
where $\bm{w}$ is the weight vector of coefficients, and $p(\cdot)$ is a sigmoid function. Here we assume that the $n$ training examples are generated independently. We thus can obtain the following log-likelihood:
\begin{eqnarray}
  \ell(\bm{w}) &=& \sum\limits_{i=1}^n \log Pr(y_i|\bm{x}_i;\bm{w})\notag\\
               &=& \sum\limits_{i=1}^n\{y_i\log p(\bm{x}_i;\bm{w})+(1-y_i)\log(1-p(\bm{x}_i;\bm{w}))\} \notag\\
               &=& \sum\limits_{i=1}^n\{y_i\bm{w}^T\bm{x}_i-\log(1+\exp(\bm{w}^T\bm{x}_i)\}.
\end{eqnarray}

\subsection{A Learning Algorithm for Logistic Regression}
Clearly, the above likelihood is a convex function. Thus we can maximize the likelihood via its gradient equations with respect to $\bm{w}$:
\begin{eqnarray}
  \nabla_{\bm{w}} \ell(\bm{w})&=& \sum\limits_{i=1}^n\{ \bm{x}_i(y_i-p(\bm{x}_i;\bm{w}))\} \notag\\
                              &=& \bm{X}^T(\bm{y}-\bm{p}) = \vec{\bm{0}},
\end{eqnarray}
where $\bm{y}=(y_1,\cdots,y_n)$ is a $n$-dimensional column vector, $\bm{X} = (\bm{x}_1,\cdots,\bm{x}_n)$ is a $n \times (p+1)$ matrix and $\bm{p} = (p(\bm{x}_1;\bm{w}),\cdots,p(\bm{x}_n;\bm{w}))$ is also a $n$-dimensional column vector. Here we adapt a Newton (or Newton-Raphson) algorithm to maximize problem (4). To this end, we first compute the Hessian matrix $\bm{H}$ of Eq. (4).
\begin{eqnarray}
  \bm{H}(\bm{w})= \nabla_{\bm{w}}^2 \ell(\bm{w}) &=& -\sum\limits_{i=1}^n\{ \bm{x}_i\bm{x}_i^Tp(\bm{x}_i;\bm{w})(1-p(\bm{x}_i;\bm{w}))\} \notag\\
                                          &=& -\bm{X}^T\bm{\Lambda}\bm{X},
\end{eqnarray}
where $\bm{\Lambda}$ is a $n\times n$ diagonal matrix with $\bm{\Lambda}_{ii} = p(\bm{x}_i;\bm{w})(1-p(\bm{x}_i;\bm{w}))$.
Starting with $\bm{w}^{old}$, we thus can obtain a Newton update rule:
\begin{equation}\label{eq-3}
\bm{w}^{new} = \bm{w}^{old} - [\bm{H}(\bm{w}^{old})]^{-1}\nabla_{\bm{w}} \ell(\bm{w})\mid_{\bm{w}=\bm{w}^{old}}.
\end{equation}
Combining Eq. (5), Eq. (6) and Eq. (7), we can rewrite the Newton update step as follows:
\begin{eqnarray}
  \bm{w}^{new} &=& \bm{w}^{old} + (\bm{X}^T\bm{\Lambda}\bm{X})^{-1}\bm{X}^T(\bm{y}-\bm{p}) \notag\\
               &=& (\bm{X}^T\bm{\Lambda}\bm{X})^{-1}\bm{X}^T\bm{\Lambda}\bm{z},
\end{eqnarray}
where $\bm{z} = \bm{X}\bm{w}^{old}+\bm{\Lambda}^{-1}(\bm{y}-\bm{p})$. We can see that the Newton update step is equivalent to solve the following linear equations with respect to $\bm{w}^{new}$ :
\begin{equation}
  (\bm{X}^T\bm{\Lambda}\bm{X})\bm{w}^{new}-\bm{X}^T\bm{\Lambda}\bm{z} = \bm{0}.
\end{equation}
Note that the diagonal matrix $\bm{\Lambda}$ and the vector $\bm{z}$ are updated using the following two formulas:
\begin{align}
  \bm{\Lambda}_{ii} & = p(\bm{x}_i;\bm{w}^{old})(1-p(\bm{x}_i;\bm{w}^{old})),\\
  \bm{z}_i & = \bm{x}_i^T \bm{w}^{old}  + \frac{y_i - p(\bm{x}_i;\bm{w}^{old})}{\bm{\Lambda}_{ii}}.
\end{align}
Since $\bm{X}^T\bm{\Lambda}\bm{X} = (\bm{\Lambda}^{1/2}\bm{X})^T(\bm{\Lambda}^{1/2}\bm{X})$ is a symmetric positive semi-definite matrix, we can obtain the optimal solution of linear system (9) via minimizing the following Quadratic Program (QP) problem:
\begin{equation}
  \bm{w}^{new} = \arg \min_{\bm{w}} \frac{1}{2} \bm{w}^T(\bm{X}^T\bm{\Lambda}\bm{X})\bm{w}-\bm{w}^T(\bm{X}^T\bm{\Lambda}\bm{z}).
\end{equation}
That is to say that each Newton update step in Eq. (7) is equivalent to minimize a QP problem. To avoid inverse matrix operation, we apply a cyclic coordinate descent method \cite{friedman2007pathwise,friedman2010regularization} to solve these QP problems. After a few steps of Newton update, we can get the optimal solution of Eq. (4). This process is also known as the iteratively re-weight least squares (IRLS) strategy \cite{P1984Iteratively,krishnapuram2005sparse}.
\subsection{A Unified Regularized Logistic Regression Framework}
Here we first introduce the Network-regularized Sparse LR (NSLR) models in a unified regularized LR framework as follows:
\begin{equation}
  \min_{\bm{w,b}}\{-\frac{1}{n}\ell(\bm{w},b) + \mathcal{R}(\bm{w})\}
\end{equation}
with
\begin{equation}
 \mathcal{R}(\bm{w}) = \lambda \{\alpha\sum_{i=1}^p |\bm{w}_i| + \frac{(1-\alpha)}{2}\bm{w}^T\bm{M}\bm{w}\},
\end{equation}
where $n$ denotes the number of samples and $\alpha \in [0,1]$. We can solve this framework via the above IRLS strategy. In other words, each Newton update step of Eq. (13) is equivalent to solving a new QP problem. For convenience, we define the coefficient vector $\bm{w} = (\bm{w}_1,\cdots,\bm{w}_p)$ and $\bm{\theta} = (\bm{w}_1,\cdots,\bm{w}_p, b)$, where $b$ is the intercept. Then, the QP problem can be defined as follows:
\begin{eqnarray}
 \mathcal{L}(\bm{w},b) = \frac{1}{2n} \bm{\theta}^T(\bm{X}^T\bm{\Lambda}\bm{X})\bm{\theta}-\frac{1}{n}\bm{\theta}^T(\bm{X}^T\bm{\Lambda}\bm{z}) + \mathcal{R}(\bm{w}),
\end{eqnarray}
where $\mathcal{R}(\bm{w})$ is a generalized network-regularized penalty. With different $\bm{M}$ in $\mathcal{R}(\bm{w})$, the penalty reduces to different cases including Lasso, Elastic net and two network-regularized ones.

\textbf{Case 1: $\bm{M} = \bm{0}$}. The penalty function Eq. (14) reduces to $\mathcal{R}(\bm{w}) = \lambda \sum\limits_{i=1}^p |\bm{w}_i|$, which is the Lasso penalty \cite{tibshirani1996regression}. We denote this case as Lasso.LR. It has been widely used in bioinformatics \cite{Shevade2003A,Cawley2006Gene,Bootkrajang2013Classification}. However, Lasso has some potential limits \cite{algamal2015regularized}. For example, it fails to select strongly correlated variable group and only selects one variable from such group and ignores the others in it.

\textbf{Case 2: $\bm{M} = \bm{I}$}. The penalty function Eq. (14) reduces to $\mathcal{R}(\bm{w}) = \lambda \sum\limits_{i=1}^p |\bm{w}_i|+ (1/2)\eta \bm{w}^T\bm{w}$, which is the so-called Elastic net penalty \cite{algamal2015regularized}. We denote this case as Elastic.LR which has adopted in \cite{algamal2015regularized}. However, to incorporate the known biological networks, a network-regularized penalty is needed in LR model.

\textbf{Case 3: $\bm{M} = \bm{L}$}, where $\bm{L}=\bm{I}-\bm{D}^{1/2}\bm{A}\bm{D}^{1/2}$ is a symmetric normalized Laplacian matrix and $\bm{A} \in \mathbb{R}^{p\times p}$ is the adjacency matrix for a given prior network (e.g., a protein interaction network). If vertex $i$ and vertex $j$ are connected, then $\bm{A}_{ij}=1$ and $\bm{A}_{ij}=0$ otherwise. The degree of vertex $i$ is defined by $d_i = \sum_{j=1}^p \bm{A}_{ij}$. Thus, $\bm{L} = (\bm{L}_{ij})$ can be re-written as follows:
\begin{eqnarray}\label{equ:02}
  \bm{L}_{ij}=
  \begin{cases}
    1, &\mbox{if}~i=j~\mbox{and}~d_i\neq 0,\cr
    -\frac{\bm{A}_{ij}}{\sqrt{d_id_j}}, &\mbox{if}~i~\mbox{and}~j~\mbox{are~ connected},\cr
    0, &\mbox{otherwise}.
  \end{cases}
\end{eqnarray}
Li and Li \cite{li2008network} applied such a network-regularized regression model for analyzing genomic data. Zhang \emph{et al.} \cite{zhang2013molecular} applied this penalty into the LR model for molecular pathway identification. We denote the Network-regularized LR model as Network.LR. They solve this model using the CVX package which was implemented for solving convex optimization problems \cite{cvx2014}. In this paper, compared with the method based on CVX, we develop a simple coordinate-wise Newton algorithm to avoid inverse matrix operation. However, the typical network-regularized penalty ignores that the pairwise variables of coefficients (linked in the prior network) may have opposite signs.

\textbf{Case 4: $\bm{M} = \bm{L^*}$}. To consider the signs of the estimated vector $\bm{w}$, we can adopt an adaptive network-based penalty $\mathcal{R}(\bm{w})$:
$$ \lambda_1 \sum\limits_{i=1}^p |\bm{w}_i| + \eta_1 \sum_{i\sim j}\bm{A}_{ij}\left(\frac{\mbox{sign}(\widehat{\bm{w}}_i)\bm{w}_i}{\sqrt{d_i}}-\frac{\mbox{sign}(\widehat{\bm{w}}_j)\bm{w}_j}{\sqrt{d_j}}\right)^2,$$
where $\lambda_1 = \lambda\alpha$, $\eta_1=\frac{\lambda(1-\alpha)}{2}$ and $\widehat{\bm{w}}$ is computed using the maximum likelihood estimation for the classical LR model. It can considered as an approximation of $|\bm{w}_j|$ using  $\mbox{sign}(\widehat{\bm{w}}_j)\bm{w}_j$. At same time, $\bm{L^*}$ can be re-written as:
 \begin{eqnarray}
  \bm{L^*}_{ij}=
  \begin{cases}
    1, &\mbox{if}~i=j~\mbox{and}~d_i\neq 0,\cr
    -\frac{\mbox{sign}(\widehat{\bm{w}}_i)\mbox{sign}(\widehat{\bm{w}}_j)A_{ij}}{\sqrt{d_id_j}}, &\mbox{if}~i~\mbox{and}~j~\mbox{are~connected,}\cr
    0, &\mbox{otherwise.}
  \end{cases}
\end{eqnarray}
Here we denote the Adaptive Network-regularized LR as AdaNet.LR.

In addition to the four cases, we also consider a novel network-regularized penalty: $\mathcal{R}(\bm{w}) = \lambda\sum_{i=1}^p |\bm{w}_i| + \eta|\bm{w}|^T\bm{M}|\bm{w}|$ \cite{min2016l0}, to consider the opposite signs of pairwise variables directly. The new penalty can eliminate the sensitivity of the typical one to the signs of the feature correlation. However, it is non-differentiable at zero point and thus the general gradient descent method cannot solve AbsNet.LR directly. In the Section 2.5, we will introduce a clever way to solve it. Note that the AdaNet.LR model can be regarded as an approximation of the AbsNet.LR model.
\subsection{A Learning Algorithm for NSLR}
Here we apply a coordinate-wise Newton algorithm to solve the NSLR framework. We first use a cyclic coordinate descent algorithm \cite{friedman2007pathwise,friedman2010regularization} to solve the QP problem in Eq. (15). Without loss of generality, we extend the $p\times p$ matrix $\bm{M}$ to a $(p+1)\times (p+1)$ matrix as follows:
\begin{eqnarray*}
\bm{M}\leftarrow
\left[
\begin{array}{c|c}
  \bm{M}  & \bm{0}  \\\hline
  \bm{0}  & 0
\end{array}
\right].
\end{eqnarray*}
Let $\lambda = n\cdot\lambda$ and $\bm{w} = (\bm{w}_1,\cdots,\bm{w}_p, b)$. We can obtain a unified form of Eq. (15):
\begin{equation}
 \mathcal{L}(\bm{w}) = \frac{1}{2} \bm{w}^T(\bm{X}^T\bm{\Lambda}\bm{X})\bm{w}-\bm{w}^T(\bm{X}^T\bm{\Lambda}\bm{z}) + \mathcal{R}(\bm{w}),
\end{equation}
where $\mathcal{R}(\bm{w}) = \lambda \{\alpha\sum_{i=1}^p |\bm{w}_i| + (1-\alpha)/2\bm{w}^T\bm{M}\bm{w}\}$.
We can easily prove that the objective function (18) is a convex one.
To minimize it, we first obtain the gradient of $\bm{w}$ as follows:
\begin{equation}
  \nabla_{\bm{w}}\mathcal{L}(\bm{w})  = (\bm{X}^T\bm{\Lambda}\bm{X})\bm{w}-\bm{X}^T\bm{\Lambda}\bm{z} + \alpha\vec{\bm{s}} + (1-\alpha)\lambda\bm{M}\bm{w},
\end{equation}
where $\vec{\bm{s}}$ is a column vector. Let $\bm{B} = \bm{X}^T\bm{\Lambda}\bm{X} + (1-\alpha)\lambda\bm{M}$, and $\bm{t} =  \bm{X}^T\bm{\Lambda}\bm{z}$. Furthermore, we can also obtain the gradient with respect to $\bm{w}_j$:
\begin{equation}
  \frac{\partial \mathcal{L}}{\partial \bm{w}_j}  = \bm{B}_{jj}\bm{w}_j + \sum_{i\neq j} \bm{B}_{ji}\bm{w}_i - \bm{t}_j + \alpha\lambda\bm{s}_j,~j = 1,\cdots,p,
\end{equation}
where $\bm{s}_j = \mbox{sign}(\bm{w}_j)$ if $\bm{w}_j \neq 0$, $\bm{s}_j = 0$ otherwise. Let
$$\frac{\partial\mathcal{L}}{\partial \bm{w}_j} = 0,$$
thus we have the following update rule for $\bm{w}_{j}$:
\begin{equation}
  \bm{w}_{j} = \mathcal{S}(\bm{t}_j-\sum_{i\neq j} \bm{B}_{ji}\bm{w}_i, \alpha\lambda)/\bm{B}_{jj},
\end{equation}
where the \textbf{soft-thresholding} function is defined as $$\mathcal{S}(a,\rho) = \mbox{sign}(a)(|a|-\rho)_+.$$

Similarly, we can also get the update rule for the bias term $b$ (note that $b = \bm{w}_{p+1}$). Given $k = p+1$, then we write the gradient of $b$:
\begin{equation}
  \frac{\partial \mathcal{L}}{\partial b}  = \bm{B}_{kk}\bm{w}_k + \sum_{i\neq k} \bm{B}_{ki}\bm{w}_i-\bm{t}_k.
\end{equation}
Let $\frac{\partial \mathcal{L}}{\partial b}=0$. It leads to the update rule for $b$:
\begin{equation}
b = (\bm{t}_k-\sum_{i\neq k} \bm{B}_{ki}\bm{w}_i)/\bm{B}_{kk}.
\end{equation}
%%%%%%%%%%%%%%%%%%%%
\begin{algorithm}[h]
\caption{NSLR learning algorithm} \label{alg:01}
\begin{algorithmic}[1]
\REQUIRE Training data $\mathcal{D}=\{\bm{X},\bm{y}\}$, a penalty matrix $\bm{M}\in \mathbb{R}^{p\times p}$, two parameters $\lambda$ and $\alpha$.
\ENSURE $\bm{w}$.
\STATE Initialize $\bm{w} = 0$
\STATE Set $\lambda = n\cdot\lambda$ where $n$ is the number of samples.
\REPEAT
\STATE Update $\bm{\Lambda}$ using Eq. (10)
\STATE Update $\bm{z}$ using Eq. (11)
\STATE Update $\bm{B} = \bm{X}^T\bm{\Lambda}\bm{X} + (1-\alpha)\lambda\bm{M}$ and $\bm{t} = \bm{X}^T\bm{\Lambda}\bm{z}$
\FOR {$j = 1$~to~$p$}
\STATE Update $\bm{w}_j$ using Eq. (21)
\ENDFOR
\STATE Update the intercept $b$ using Eq. (23)
\STATE Compute the criteria $J = -(1/n)\ell(\bm{w},b) + \mathcal{R}(\bm{w})$ for testing convergence
\UNTIL The objective function $J$ converges a minimum
\RETURN $\bm{w}$
\end{algorithmic}
\end{algorithm}
Briefly, here we employ the iteratively re-weight least squares (IRLS) strategy to solve the unified regularized LR framework. In each iteration, we first update Eq. (10) and Eq. (11) to get a new constrained QP problem (18). Then we apply a cycle coordinate descent algorithm to solve it. This process is repeated until convergence. To summarize, we propose the following Algorithm 1.
\subsection{A Novel Network-regularized LR}
In the subsection, we focus on a novel Network-regularized Logistic Regression model with absolute operation (AbsNet.LR) as follows:
\begin{equation}
  \min_{\bm{w,b}}\{-\frac{1}{n}\ell(\bm{w},b) + \lambda\sum_{i=1}^p |w_i| + \eta|\bm{w}|^T\bm{L}|\bm{w}|\},
\end{equation}
where $\ell(\bm{w},b) = \sum\limits_{i=1}^n\{y_i\bm{w}^T\bm{x}_i-\log(1+\exp(\bm{w}^T\bm{x}_i)\}$ and $|\bm{w}| = (\bm{w}_1, \bm{w}_2,\cdots, \bm{w}_p)^T$. As we have discussed that AdaNet.LR model can be considered as an approximation of it (see Section 2.3 for Case 4). AdaNet.LR applies a convex penalty to replace the one in Eq. (24) enabling it can be solved by NSLR algorithm (Algorithm 1). In the subsection we employ a coordinate-wise Newton algorithm to solve it directly. For each Newton update step, we focus on the following optimization problem:
\begin{equation}
 \mathcal{L}(\bm{w},b) = \frac{1}{2n} \bm{w}^T(\bm{X}^T\bm{\Lambda}\bm{X})\bm{w}-\frac{1}{n}\bm{w}^T(\bm{X}^T\bm{\Lambda}\bm{z}) + \mathcal{R}(\bm{w}),
\end{equation}
where $\mathcal{R}(\bm{w}) = \lambda \{\alpha\sum_{i=1}^p |\bm{w}_i| + (1-\alpha)/2|\bm{w}|^T\bm{L}|\bm{w}|\}$. The coordinate-wise descent strategy does not work for some penalties (e.g., fused lasso) \cite{ hastie2015statistical}. However, Hastie \emph{et~al.} \cite{hastie2015statistical} (see page 112) find the penalty $\mathcal{R}(\bm{w})$ shows a good property: condition regularity, implying that if the iteration moving along all coordinate directions fails to enable the objective function decrease, then it achieves the minimum \cite{hastie2015statistical,Tseng2001Convergence}. Thus we can minimize Eq. (25) via a cycle coordinate descent method. For convenience, let $\bm{B} = \bm{X}^T\bm{\Lambda}\bm{X}$, $\bm{t} = \bm{X}^T\bm{\Lambda}\bm{z}$ and $\eta = \lambda(1-\alpha)/2$, we can rewrite the subproblem as follows:
\begin{eqnarray}
  \mathcal{L}_k &=& \frac{1}{2n}(\bm{B}_{kk}\bm{w}^2_k + 2\sum_{j\neq k}{\bm{B}_{kj}\bm{w}_j\bm{w}_k} + \sum_{j\neq k}\sum_{i\neq k}{\bm{B}_{ji}\bm{w}_j\bm{w}_i}) \notag\\
                & & - \frac{1}{n}(\bm{w}_k\bm{t}_k + \sum_{j\neq k}{\bm{w}_j\bm{t}_j}) + \lambda(|\bm{w}_k|+\sum_{j\neq k}{|\bm{w}_j|})\notag\\
                & & + \eta\bm{L}_{kk}\bm{w}^2_k + 2\eta\sum_{j\neq k}{\bm{L}_{kj}|\bm{w}_j||\bm{w}_k}| \notag\\
                & & + \eta\sum_{j\neq k}\sum_{i\neq k}{\bm{L}_{ji}|\bm{w}_j||\bm{w}_i|}.
\end{eqnarray}
Then we have the subgradient equation of $\bm{w}_k$:
\begin{eqnarray}
\frac{\partial \mathcal{L}}{\partial \bm{w}_k} &=& (1/n)(\bm{B}_{kk}\bm{w}_k+\sum_{j\neq k}{\bm{B}_{kj}\bm{w}_j})-(1/n)\bm{t}_k + \lambda \bm{s}_k \notag\\
                                               & & +2\eta(\bm{L}_{kk}\bm{w}_k + \bm{s}_k \sum_{j\neq k}{\bm{L}_{kj}|\bm{w}_j}|).
\end{eqnarray}
Let $ \frac{\partial \mathcal{L}}{\partial \bm{w}_k}=0$ and $\tau = \bm{B}_{kk}+2n\eta\bm{L}_{kk}$. We obtain the following update rule:
\begin{equation}
\bm{w}_k = \mathcal{S} (\bm{t}_k-\sum_{j\neq k}{\bm{B}_{kj}\bm{w}_j},n\lambda+2n\eta\sum_{j\neq k}{\bm{L}_{kj}|\bm{w}_j}|)/\tau.
\end{equation}
Similarly, we can also get the update rule for the bias term $b$. The sub-gradient is given by
\begin{equation}
  \frac{\partial \mathcal{L}}{\partial b}  = \bm{B}_{kk}\bm{w}_k + \sum_{i\neq k} \bm{B}_{ki}\bm{w}_i-\bm{t}_k,
\end{equation}
where $k = p+1$, and $b = \bm{w}_k$. Let $\frac{\partial \mathcal{L}}{\partial b}=0$. It leads to the update rule:
\begin{equation}
b = (\bm{t}_k-\sum_{i\neq k} \bm{B}_{ki}\bm{w}_i)/\bm{B}_{kk}.
\end{equation}
To summarize, we propose the Algorithm 2 to solve the AbsNet.LR model.
\begin{algorithm}[h]
\caption{AbsNet.LR learning algorithm} \label{alg:02}
\begin{algorithmic}[1]
\REQUIRE Training data $\mathcal{D}=\{\bm{X},\bm{y}\}$, a normalized Laplacian matrix $\bm{L}\in \mathbb{R}^{p\times p}$, two parameters $\lambda$ and $\alpha$.
\ENSURE $\bm{w}$.
\STATE Initialize $\bm{w} = 0$
\STATE Set $\lambda = n\cdot\lambda$, where $n$ is the number of samples.
\REPEAT
\STATE Update $\bm{\Lambda}$ using Eq. (10)
\STATE Update $\bm{z}$ using Eq. (11)
\STATE Update $\bm{B} = \bm{X}^T\bm{\Lambda}\bm{X}$ and $\bm{t} =  \bm{X}^T\bm{\Lambda}\bm{z}$
\FOR {$k = 1$~to~$p$}
\STATE  Update $\bm{w}_k$ using Eq. (28)
\ENDFOR
\STATE Update the intercept $b$ using Eq. (30)
\STATE Compute the criteria $J = -(1/n)\ell(\bm{w},b) + \mathcal{R}(\bm{w})$ for testing convergence
\UNTIL The objective function $J$ converges to a mimimum
\RETURN $\bm{w}$
\end{algorithmic}
\end{algorithm}

\textbf{Tuning parameter selection.} Selecting the regularized parameters $\{\lambda,\alpha\}$ for NSLR and AbsNet.LR is a very important task. Here these parameters are selected via maximizing the index -- Area Under Curve (AUC). Suppose there are $n_1$ positive class samples and $n_2$ negative class samples in a given dataset. Given a binary classifier, $\{s_1,\cdots,s_{n_1}\}$ are the scores for the positive points and $\{t_1,\cdots,t_{n_2}\}$ are the scores for the negative points. The AUC of this classifier is calculated using the following formula:
\begin{equation}
  \mbox{AUC} = \frac{1}{{n_1n_2}}\sum\limits_{i=1}^{n_1} \sum\limits_{j=1}^{n_2}\big(I(s_i>t_j)+ \frac{1}{2}I(s_i=t_j) \big),
\end{equation}
where $I(s_i>t_j)$ is 1, if $s_i>t_j$, and 0 otherwise. However, sometimes the parameter selection may be unnecessary in a real application. For example, we may choose suitable parameters to obtain a solution for the desired degree of sparsity.
\section{Synthetic Data Experiments}
To compare the performance of different regularized LR models, we first generate a sparse coefficient vector $\bm{w}$ with $p=100$ and a bias term $b$ as follows:
$$\bm{w} = [sample(\mathcal{N}(0,1),40),rep(0,60)],~b=0,$$
where $sample(\mathcal{N}(0,1),40)$ denotes a vector of length 40, whose elements are sampled from a standard normal distribution, and $rep(0,60)$ denotes a vector of length 60, whose entries are zero. To generate an expression matrix $\bm{X} \in \mathbb{R}^{500\times100} $, we define a covariance matrix of the variables $\bm{\Sigma} \in \mathbb{R}^{100\times100}$, where $\bm{\Sigma}_{ij} = 0.6$ when $1\leq i \neq j \leq 40$, $\bm{\Sigma}_{ii} = 1$ when $i=1,\cdots,100$ and the others entries are zero. It can ensure that the 40 feature vectors are strong correlated in $\bm{X}$. We generate the $\bm{X}$ using the function ``mvrnorm'' in the MASS R package with parameters $mu = \bm{0}$ and $sigma = \bm{\Sigma}$. Furthermore, we also compute the binary response $\bm{y}=(\bm{y}_1,\cdots,\bm{y}_{500})$ based on a Bernoulli distribution with the following formula:
$$\bm{y}_i = I(p(\bm{x}_i)\geq0.5),~ p(\bm{x}_i) = \frac{1}{1+\exp(-\bm{w}^T\bm{x}_i-b)},$$
where $\bm{x}_i$ denotes the $i$-th column of $\bm{X}^T$. Finally, we also generate a prior network $\bm{A}\in \mathbb{R}^{100\times100}$, whose node pairs among all the first 40 nodes are connected with probability $p_{11}=0.3$, and the remaining ones are connected with probability $p_{12}=0.1$. Note we set the observed matrix $\bm{X}^* = \bm{X}+ \gamma \bm{\epsilon}$, where the elements of $\bm{\epsilon}$ are randomly sampled from a standard normal distribution, and $\gamma=3$ is used to control the signal-to-noise ratio. The observed $\bm{X}^*$ contains 500 samples with 100 features. We randomly select 300 samples as the training samples and the remaining samples as the testing samples. We test all regularized LR models on the synthetic training data using 5-fold cross-validation strategy to select the optimal parameters. We repeat the simulations over 50 times. We then compute the average AUCs of 50 experiments about classification on the testing synthetic data. To compare the performance on variable selection, we also calculate the average sensitivity/specificity scores with respect to the selected variables (nonzero of $\bm{w}$), and the average numbers of edges of the selected variables in the prior network.

Here we define the \textbf{sensitivity} as the percentage of true non-zero entries discovered, and the \textbf{specificity} as the percentage of true zero entries discovered, respectively. The overall results are shown in Table 1. Generally, these network-based regularized LR models (especially AbsNet.LR) are superior to other algorithms with respect to AUCs. Moreover, AbsNet.LR obtains the relatively higher sensitivity in the selected variables, and relatively larger number of edges of them in the prior network compared to others.
\begin{table}[ht]
\centering
\caption{Results on the synthetic data.}
\resizebox{\columnwidth}{!}{
\begin{tabular}{l|l||cccc}
  \hline
              &      &$\bm{w}$  & no. of  & $\bm{w}$   &$\bm{w}$    \\
  methods     &AUC   &nonzero   & edges   &sensitivity &specificity\\
  \hline
  LR          &0.733  & 100   &450.00    &1.00   &0.00\\
  Lasso.LR    &0.792  & 45.84  &134.22     &0.67   &0.68\\
  Elastic.LR  &0.806  & 52.76  &176.62     &0.79   &0.64\\
  Network.LR  &0.828  & 54.64  &234.42     &0.97   &0.74\\
  AdaNet.LR   &0.823  & 53.20  &196.40     &0.88   &0.70\\
  AbsNet.LR   &\textbf{0.830}  & 63.50   &\textbf{281.78}   &\textbf{0.98}  &0.60\\
  \hline
\end{tabular}
}
\end{table}
\section{Application to GBM data}
\subsection{The GBM Data}
We download the level 3 gene expression data (Broad Institute HT-HG-U133A platform) and clinical data of GBM from the TCGA database \cite{Benz2013The}. We employ two alternative methods to impute the missing data in gene expression profiling: (1) the $k$-nearest neighbor algorithm (knn) method implemented as the ``impute" R package; (2) the mean of considered genes. We find that the two methods only have little effect on the final results. Thus, we simply adopt the second method to impute the missing data. We obtain the gene expression data of 1,2042 genes across 525 patients. Furthermore, we standardize the expression of each gene across all samples using the ``scale'' function in R. We also download a protein interaction network (PPI) data from Pathway Commons \cite{Cerami2010Pathway}. Finally, we obtain a gene expression data with 1,0963 genes and a PPI network with 24,8403 edges. Here our goal of this application is to predict the survival risk of GBM patients (Fig. 1A). We first dichotomize the survival time of patients into a binary outcome through a designated cutoff. Here we consider one year as the cutoff time to balance the number of positive and negative samples (Fig. 1B and Table 2).

\begin{table}[ht]
\centering
\caption{Description of the GBM data. $^\star$denotes the number of patients which are alive/death within one year}
\resizebox{\columnwidth}{!}{
\begin{tabular}{lccccc}
  \hline
   data &  \#genes & \#samples & \#alive$^\star$ & \#death$^\star$ & \#censoring  \\
  \hline
   GBM  & 1,0963       & 525      & 236             & 204 & 115    \\

  \hline
\end{tabular}
}
\end{table}

We apply the univariate Cox proportional hazard model \cite{Cox1972} to assess and filter the 1,0963 genes of GBM gene expression data (Table 2). Finally, we obtain 2001 genes with $P<0.05$ for further analysis. Only those genes are used in all the regularized LR models.
\subsection{Results of the GBM Data}
We first randomly select 2/3 samples ($n$ = 350) as the training samples and 1/3 samples ($n$ = 175) as the testing samples (Fig. 1B). We consider $\lambda \in \{0.001, 0.01, 0.05, 0.1, 0.2, 0.3, 0.4\}$ and $\alpha \in \{0.05, 0.1, 0.2, 0.5, 0.7, 0.9\}$ to form a total of 42 pair parameter sets. We first learn the different regularized LR methods on the GBM training data using 5-fold cross-validation. Then we test all the methods on the GBM testing data. We find the regularized LR models are superior to the typical LR model (Table 3). Generally, Lasso.LR is inferior to other regularized methods (Elastic.LR, Network.LR, AdaNet.LR and AbsNet.LR) whose results are relatively consistent. However, AbsNet.LR, AdaNet.LR and Network.LR identify a gene set with more edges via integrating the prior network data. All these results imply the importance of integrating the prior protein interaction network data to improve the prediction accuracy and biological interpretation. Therefore, we only focus on the result analysis of AbsNet.LR which obtains AUC $= 0.6627$ with $\lambda = 0.3$ and $\alpha = 0.1$, and a gene set with 157 genes.
\begin{table}[ht]
\centering
\caption{Summary of 42 trials in the independent test data. ``avg.no.edge'' denotes the average number of the identified genes. Similarly, ``avg.no.edge'' denotes the average number of edges between the identified genes. ``avg.AUC'' denotes the average AUC of all the 42 trials. ``max.AUC'' denotes the maximal AUC of 42 trials.}
\resizebox{\columnwidth}{!}{
\begin{tabular}{l|l|l|c|c}
  \hline
  methods & avg.no.gene & avg.no.edge & avg.AUC & max.AUC \\
  \hline
  LR          & 2001     & 6341      & 0.5787 & 0.5787 \\
  Lasso.LR    & 152.2619 & 45.0714   & 0.5673 & 0.6182 \\
  Elastic.LR  & 324.1429 & 296.9762  & 0.5842 & 0.6659 \\
  \hline
  Network.LR  & 325.6905 & 328.1905  & 0.5832 & 0.6660 \\
  AdaNet.LR   & 331.7857 & 448.9524  & 0.5823 & 0.6659 \\
  AbsNet.LR   & 410.6429 & 1013.6190 & 0.5860 & 0.6627 \\
  \hline
\end{tabular}
}
\end{table}

\begin{figure*}[htbp]
  \centering
  \includegraphics[width = 0.65\linewidth]{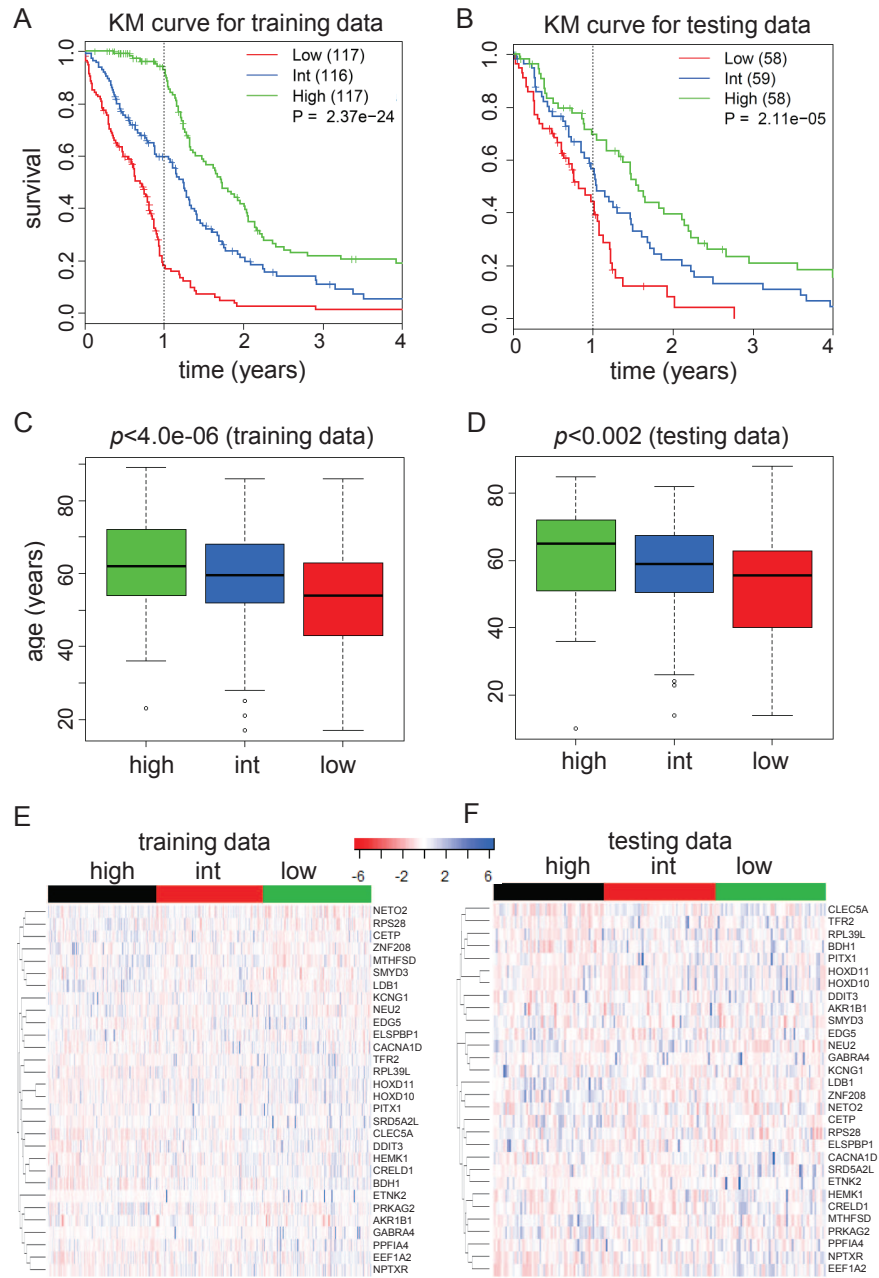}\\
  \caption{Performance evaluation in the training and testing data. We divide the GBM patients into different groups according to their estimated LR scores. (A) and (B) Kaplan-Meier (KM) survival curves for for low-, intermediate (int)-, and high-risk groups. $P$-values are computed using the log-rank test. (C) and (D) Age distribution in the three GBM patient groups with $p$-value $<$4.0e-06 for training data and $p$-value $<$ 0.002 for testing data (using an analysis of variance model ). (E) and (F) The expression heat maps of the training and testing data based on the top 30 genes.}\label{fig:01}
\end{figure*}

To evaluate the biological relevance of the identified gene set, we apply the DAVID online web tool \cite{huang2009systematic} to perform the Gene Ontology (GO) biological processes enrichment analysis. The GO biological processes with $p$-values $<$ 0.05 are selected as significant ones. We find the genes is related to some important ones including positive regulation of kinase activity, synaptic transmission glutamatergic and inositol metabolic process. The identified biological processes are consistent with recent literature reports. For example `the positive regulation of protein kinase activity' process is well-known to be related to GBM and its activation is a key mechanism for GBM development \cite{azoitei2011protein}.

Furthermore, we extract the top 30 genes corresponding to the largest absolute values of the estimated coefficients (by the AbsNet.LR). Based on the top 30 genes, we re-train the typical LR model in the training data. In the independent testing data, LR obtain an AUC of 0.6727. However, on average, only AUC of 0.55 are obtained with randomly select 30 genes. Next, based on the built LR model, we re-test all the training data and all testing data (Fig. 1B) and based on the predicted LR scores computed via $Pr(y=1|\bm{x};\bm{w})$, we divide all the training and testing samples into three groups, called low-, intermediate- and high-risk ones, respectively (Fig. 2A and B). We can also see the expression heat maps of the top 30 genes with the three groups (Fig. 2E and 2F). We find that some genes in the specific group are high expressed, while some are low expressed in the other group. Interestingly, we also find that these top genes form a set of sub-networks (e.g., Fig. 3A and B), which are sugested to be related with GBM  \cite{Viotti2013Glioma,Biasoli2014Glioblastoma,Hu2013Targeting,Calzolari2010Transferrin}. Furthermore, we study whether the defined groups have significant relationships with some other clinical variables (e.g., age). We find that the patients of high-risk group are old than those in low and intermediate-risk group on both training and testing data (Fig. 2C and D). All these results show that the identified gene set is related to the survival outcome of GBM patients.
\begin{figure}[htbp]
  \centering
  \includegraphics[width = 1\linewidth]{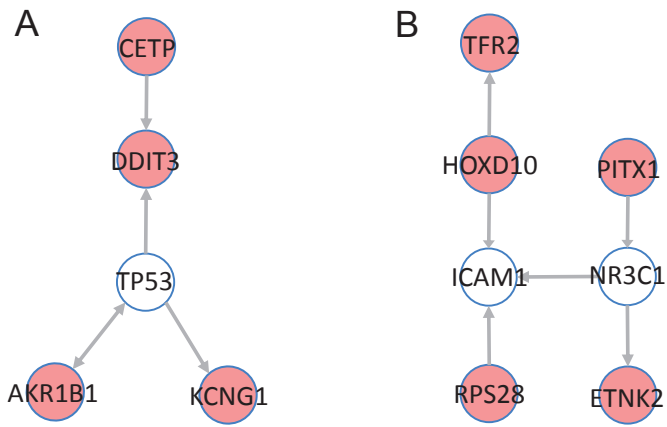}\\
  \caption{Two known pathway sub-networks of the identified top genes. Three genes (\emph{TP53}, \emph{ICAM1} and \emph{NR3C1}) are ``linked genes¡± included by using Ingenuity Pathway Analysis (IPA, http://www.ingenuity.com) software. %(A) is a sub-network regularized by TP53 which is reported to be related with GBM \cite{Viotti2013Glioma,Biasoli2014Glioblastoma} and these genes (TFR2, HOXD10, RPS28, PITX1, ETNK2) included in the sub-network (B) are also reported to be related with GBM cancer \cite{Hu2013Targeting,Calzolari2010Transferrin,Kim2016ccmGDB}.
  }\label{fig:03}
\end{figure}
\section{Application to Lung cancer data}
\subsection{Non-Small Cell Lung Cancer Expression Data}
We download two subtypes of non-small cell lung cancer gene expression datasets from TCGA \cite{Hoadley2014Multiplatform}, including Lung adenocarcinoma (LUAD) with 350 samples and Lung squamous-cell carcinoma (LUSC) with 315 samples. Here we consider the level 3 expression data of TCGA which is quantified at the gene and transcript levels using RNA-Seq data by an expectation maximization method \cite{Chuong2008What}. Then the expression values are $\log_2$-transformed. We first process the LUAD and LUSC gene expression data to keep the genes that are appeared in the KEGG pathway database from the Molecular Signatures Database (MSigDB) \cite{subramanian2005gene}. Finally, we obtain 3230 genes related with 151 KEGG pathways. Here we consider each KEGG pathway as a fully connected sub-network to build a prior network with 143438 edges.
\subsection{Results of the Lung Cancer Data}
Here we apply all the methods to the lung cancer data to identify subtype-specific biomarkers. Compared to the integration of PPI network in the GBM data, we focus on the KEGG pathway as the prior information. We randomly select all the samples of 2/3 as the training samples and remaining ones are as the independent test samples. We show the results from all the regularized LR models in the independent training set (Table 4). In total, 69 genes are selected by the AbsNet.LR. All the genes identified by other regularized LR models (Lasso.LR, Elastic.LR, Network.LR and AdaNet.LR) are included in the 69 genes. The network-regularized LR models give similar results, which are superior to that of typical LR and Lasso.LR models. In addition, compared to the other network-regularized LR models, AbsNet.LR identifies a gene set with more edges in the KEGG pathway network.
\begin{table}[ht]
\centering
\caption{Results of Lung cancer data.}
\resizebox{\columnwidth}{!}{
\begin{tabular}{l|c|c|c}
  \hline
  methods & AUC & no. of genes & no. of edges\\
  \hline
  LR          & 0.7540 & 3230   & --    \\
  Lasso.LR    & 0.9784 & 48 & 53 \\
  Elastic.LR  & 0.9789 & 57 & 62 \\
  \hline
  Network.LR  & 0.9789 & 59 & 67 \\
  AdaNet.LR   & 0.9789 & 57 & 62 \\
  AbsNet.LR   & 0.9794 & 69 & 104\\
  \hline
\end{tabular}
}
\end{table}

Furthermore, we evaluate the biological relevance of the identified 69 genes using the DAVID online web tool \cite{huang2009systematic} and find several significantly enriched KEGG pathways relating to lung cancer, including metabolism of xenobiotics by cytochrome P450 (GSTA2, CYP3A5, CYP2F1, CYP3A7, CYP2C9, CYP2C8, UGT2A1) \cite{Hukkanen2002Expression}, linoleic acid metabolism (CYP3A5, CYP3A7, CYP2C9, AKR1B10, CYP2C8) \cite{Liu2014Serum} and retinol metabolism (CYP3A5, CYP3A7, CYP2C9, CYP2C8, UGT2A1, RPE65) \cite{Kuznetsova2016Abnormal} and so on.
\section{Computing Platform and Running Time}
All scripts were run on a desktop personal computer with an Inter(R) Core(TM) i7-4770 CPU@ 3.4 GHz and 16 GB memory running Windows OS and R 3.2.5. The code is available at http://page.amss.ac.cn/shihua.zhang/. For the synthetic data experiments, the memory used is about 140 MB and the running time is about 0.7 hours. For the application to GBM data, the memory used is about 150 MB and the running time is about 0.8 hours. For the application to lung cancer data, the memory used is about 160 MB and the running time is about 1 hour.
\section{Conclusion}
In this paper, we first introduce the typical network-regularized LR models with others in a unified framework. Although the typical network-regularized LR model can incorporate such prior information to get better biological interpretability, it fails to focus on the opposite effect of variables with different signs. To solve this limitation, we adopt a novel network-regularized penalty $\mathcal{R} = \lambda \|\bm{w}\|_1 + \eta|\bm{w}|^T\bm{L}|\bm{w}|$ into LR model (denoted as AbsNet.LR). However, the novel penalty is not differentiable at the zero point, enforcing it is hard to be solved using the typical gradient descent method. To this end, we first adapt the adaptive network-regularized penalty (note that it is convex) to approximate this novel network-regularized penalty and develop an adaptive network-regularized LR (AdaNet.LR) model which can be solved easily using the NSLR algorithm (Algorithm 1). We further find that the novel network-regularized penalty has a good property -- condition regularity \cite{hastie2015statistical,Tseng2001Convergence}, which inspires us to solve it via a cycle coordinate-wise Newton algorithm efficiently. We note that the binary LR-based classification models can be easily extended to the multiclass or multinomial classification problem \cite{krishnapuram2005sparse,Sch2006Sparse}.

Applications to the synthetic data show that the present methods are more effective compared to the typical ones. We also apply them to the GBM expression and a protein interaction network data to predict mortality probability of patients within one year. In addition, we apply them to a lung cancer expression data of two subtypes to identify subtype-specific biomarkers. We find a gene set with 69 genes which are tightly connected in the prior network. Functional enrichment analysis of these genes discovers a few KEGG pathways relating to the lung cancer clearly.
\section*{Acknowledgment}
Shihua Zhang and Juan Liu are the corresponding authors of this paper. Wenwen Min would like to thank the support of National Center for Mathematics and Interdisciplinary Sciences, Academy of Mathematics and Systems Science, CAS during his visit. This work was supported by the National Science Foundation of China [61379092, 61422309, 61621003], the Strategic Priority Research Program of the Chinese Academy of Sciences (CAS) (XDB13040600), the Outstanding Young Scientist Program of CAS, CAS Frontier Science Research Key Project -- Top Young Scientist (No. QYZDB-SSW-SYS008) and the Key Laboratory of Random Complex Structures and Data Science, CAS.

%\\
\bibliographystyle{IEEEtran}
\bibliography{References}

\balance
\end{document}